\documentclass[prb,showpacs,twocolumn]{revtex4}
\usepackage{graphics}
\begin{document}
\title{Experimental observation of charge ordering in nanocrystalline 
Pr$_{0.65}$Ca$_{0.35}$MnO$_{3}$}
\author{Anis Biswas}
\email{anis.biswas@saha.ac.in}
\author{I. Das}
\affiliation{Saha Institute of Nuclear Physics,1/AF,Bidhannangar,
Kolkata 7000 064,India}
\begin{abstract} Observation of charge ordering in single crystalline
 and bulk polycrystalline systems of various rare-earth based manganites
 is well documented. However, there is hardly any manifestation of the
 same when the grain size is reduced  to nanoscale. We have observed
 charge ordering in case of nanocrystalline  
Pr$_{0.65}$Ca$_{0.35}$MnO$_{3}$ of average particle size $40$ nm.
This phenomenon is attributed to the primary role played by the martensitic
 character of the charge order transition in the material.
\end{abstract}
\maketitle
\section{introduction}
\noindent
The phenomenon charge ordering (CO) in perovskite manganites with general 
formula R$_{1-x}$B$_{x}$MnO$_{3}$ (where, R is rare-earth, B is bivalent ion)  
has become a subject of intense research \cite{collosal,nagaev}. This 
intriguing phenomenon is associated with strong interplay between charge, 
lattice and orbital degrees of freedom and mainly observed for some 
commensurate fraction of carrier concentration like x = $1/2, 2/3,4/5$ etc. 
However, the formation of CO state is also possible for other incommensurate 
values of carrier concentration.  
CO state can be destabilized by magnetic field and gives rise to a large
 negative magnetoresistance (MR).
In literature, there are many studies regarding CO in polycrystalline and
singlecrystalline bulk form of the sample \cite{collosal,nagaev}.
The primary ingredients behind the formation of CO state are emphasized
 as the competition between double exchange (DE) and super exchange(SE) among
 the core spins of manganese and the coulomb interaction between electrons
 of different orbitals of the same manganese-site \cite{brink1,brink2}.
The Charge Order transition is accompanied by structural transition.
This structural effect has been considered as the secondary effect of CO.
More recently, the polarized optical studies have revealed that CO is a 
martensitic-like transformation\cite{podzorov}. 
The martensitic strain is another important
 factor in case of CO \cite{podzorov,mar1,mar2}.       
The phenomenon CO is not much explored in case of nanocrystalline form of the 
samples. The issue is still not clear that which factor plays the 
dominant role in CO especially, when the particle size of the system is 
reduced.
In this report our primary objective is to address this issue by 
studying  the phenomenon in  nanocrystalline material. 

The Ca doped Pr- manganites (Pr$_{1-x}$Ca$_{x}$MnO$_{3}$) with the doping
 concentration $0.3$$\le$x$\le$$0.5$ in bulk form show same generic behavior 
in their phase diagram \cite{urushihara}. They remain at insulating 
state in all temperatures in absence of any external perturbation. 
Charge Order transition occurs below a certain temperature (T$_{CO}$)
\cite{urushihara}. The antiferromagnetic transition does not
 coincide with CO. The antiferromagnetic transition temperature, T$_{N}$, is 
always less than T$_{CO}$ \cite{urushihara}. For the samples having 
Ca-concentration away from 
commensurate value `$0.5$', the antiferromagnetic state is transformed to a 
canted antiferromagnetic state at low temperature well below T$_{N}$ 
\cite{urushihara}. 

We have performed a detailed experimental study
 on the nanocrystalline Pr$_{0.65}$Ca$_{0.35}$MnO$_{3}$ of average particle
 size $\sim$ $40$ nm and CO state has been observed.
In case of the bulk Pr$_{0.65}$Ca$_{0.35}$MnO$_{3}$,T$_{CO}$ and T$_{N}$ are 
$225$ K and $175$ K respectively \cite{bulk}.The canted antiferromagnetic 
structure has been stabilized below $\sim$ 100 K \cite{bulk}. 
 
\section{Sample Preparation and Characterization}
The nanoparticles of Pr$_{0.65}$Ca$_{0.35}$MnO$_{3}$ have been prepared by 
sol-gel technique. At the end of the process, the gel is decomposed at about 
$100$$^{o}$ C and a porous black powder has been obtained. The powder is given 
heat treatment at $1000$$^{o}$ C for $6$ hour to get nanocrystalline sample.
The x-ray diffraction study has confirmed the single phase nature of the
 sample with orthorhombic crystal structure (pbnm symmetry) similar to the bulk
 sample \cite{physica}. The values of the lattice constants (a = $5.420$ \AA, 
 b = $5.449$ \AA and c = $7.660$ \AA) of the sample agree quite well 
with literature values for the bulk sample \cite{physica}.
Transmission Electron Microscopy (TEM) study has revealed that
 the average particle size is $\sim$ $40$ nm.
 One typical electron micrograph of the sample has been shown in Fig. 1. 
The histogram showing particle size distribution as well as the High-resolution
 Transmission Electron Microscopy (HRTEM) image of the inside portion of a 
particle has been depicted in the insets of Fig.1. The HRTEM picture 
(inset [B], Fig. 1) exhibits clear lattice- planes implying  good
 crystallinity inside the particle.  
The crystallite size of the sample is calculated from the full width of the
 half maximum (FWHM) of x-ray diffraction peaks using Scherre formula 
\cite{scherrer}.The correction due to instrumental broadening has been
 taken into account \cite{warrer, cryst}. The calculated average crystallite 
size is $\sim$ $36$ nm. 

\begin{figure}
\resizebox{6cm}{6cm}
{\includegraphics{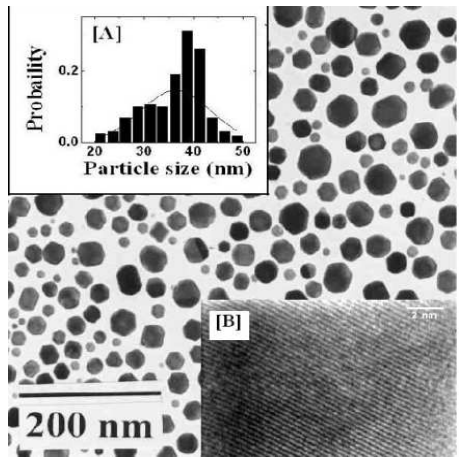}}
\caption{Transmission Electron Micrograph of 
nanocrystalline sample of Pr$_{0.65}$Ca$_{0.35}$MnO$_{3}$. Inset:[A]Histogram
 showing particle size distribution [B] High-Resolution Transmission
 Electron Microscopy (HRTEM) image of inside portion of a particle
 showing lattice resolved planes.}
\end{figure}

\section{Experimental Results and Discussions}
Resistivity measurement has been performed in usual four probe method in 
presence as well as in absence of magnetic field. In absence of the
 external magnetic field, resistivity increases with 
the decrease of temperature in the entire range. The large
 resistance below $\sim$ $80$ K exceeds the maximum measurable range of our
 experimental set-up. The signature of transition is not visible. The 
application of magnetic field has drastic effect on the transport (Fig. 2). 
 There is a tendency of insulator to metal 
transition below $\sim$ $100$ K in the presence of $40$ kOe magnetic field. 
However, complete insulator to metal transition is achieved in the presence of 
 $60$ kOe magnetic field.
\begin{figure}
\resizebox{8cm}{6cm}
{\includegraphics{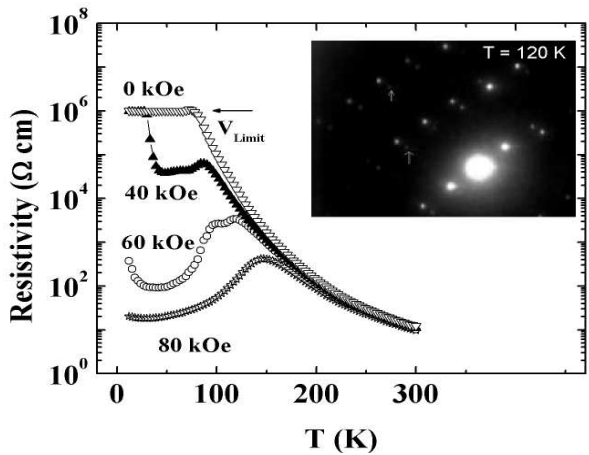}}
\caption{Resistivity as a function of temperature for
nanocrystalline sample of Pr$_{0.65}$Ca$_{0.35}$MnO$_{3}$ of average particle
size $50$ nm in presence as well as in absence of magnetic field. Inset:
 Electron Diffraction Pattern of Selected Area (SAED) of the sample at 
$120$ K. The super lattice spots indicated by arrows are signatures
 of Charge Order transition.}
\end{figure}
 The temperature at which the magnetic field induced 
insulator to metal transition occurs (T$_{IM}$) shifts to higher temperature 
with the increase of magnetic field. 
The transport measurements on the nanocrystalline samples indicates that the
temperature 
dependence of resistivity (Fig. 2) in 
presence as well as in absence of magnetic field is almost similar to the 
bulk sample \cite{prca1} except the invisibility of any sharp transition
 in absence of magnetic field. For the 
bulk sample \cite{prca1}, the increase of resistivity at low temperature is
 due to CO. Such a high value of resistivity at low temperature for 
nanocrystalline 
sample is expected for  the existence of the CO state.The 
switching over of resistivity from insulating  to metallic state in presence
 of magnetic field may  be possible due to the melting of the CO state by 
magnetic field for this kind of system. The shifting of T$_{IM}$ to the high 
temperature
 region with the increase of magnetic field is also observed for the
 bulk sample \cite{prca1}.                                                      

The existence of CO state at low temperature has been further verified
 by electron diffraction study.In the  Electron Diffraction pattern at low 
temperature, superlattice  spots have been observed (Inset, Fig. 2) which exist as a result of CO transition as seen in various earlier experimental studies
\cite{chen,edprl,liu}.

The transport measurements along with electron diffraction study clearly 
indicate the existence of the CO state in case of the present 
nanocrystalline sample. 
The grain boundaries of the nanoparticles are in disordered state and have
strong extrinsic effect on transport. The intrinsic 
character of transition can be affected by this extrinsic effect. 
To get better understanding of transition, heat capacity measurement has
 been performed. 
\begin{figure}
\resizebox{8cm}{8cm}
{\includegraphics{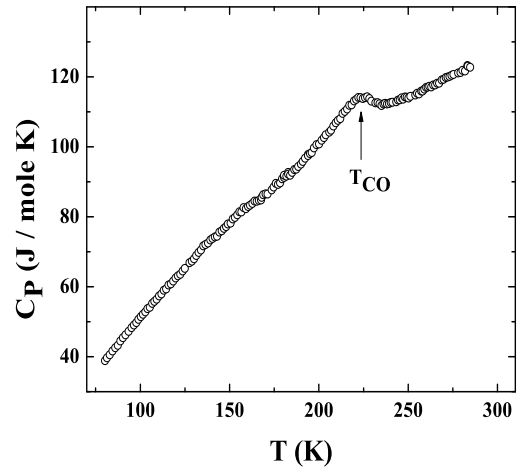}}
\caption{Specific heat as a function of temperature for
nanocrystalline sample of Pr$_{0.65}$Ca$_{0.35}$MnO$_{3}$ of average particle
 size $40$ nm
in absence of the magnetic field. }
\end{figure}
The temperature dependence of specific heat (Fig. 3) indicates a clear peak at 
$\sim$ $225$ K. The temperature at which
the peak is observed is almost same as the reported Charge Order transition
 temperature (T$_{CO}$)  of the bulk form
 of the sample \cite{bulk}. The peak in specific heat data 
in case of the nanocrystalline sample indicates CO transition occurs at almost
 same temperature as that of the bulk sample.

A commercial SQUID magnetometer has been employed to study magnetic properties 
of the sample.
The temperature dependence of zero field cooled (ZFC) DC susceptibility in 
 the presence of $1$ kOe magnetic field has been shown in Fig. 4. There exist
 a peak around $\sim$ $225$ K (inset [A], Fig. 4), close to the reported 
T$_{CO}$ of the bulk sample \cite{bulk}.
 The position of the peak coincides with the temperature
 at which peak is obtained in the temperature dependence of specific heat.
The existence of the peak is another indication of the charge order transition 
for the present nanoparticles. There is no clear signature of 
antiferromagnetic transition unlike the bulk sample.
The susceptibility obeys Curie-Weiss behavior for $T > T_{CO}$
 with paramagnetic Courie temperature ($\theta_{p}$) $\sim$
 $157$ K (inset [B], Fig. 4). 
The susceptibility starts 
to increase sharply below $\sim$ $100$ K (Fig 4), 
which seems to be associated to the onset of canted antiferromagnetic
 (CAF) structure as observed in case of the bulk sample \cite{bulk}. 
\begin{figure}
\resizebox{8cm}{8cm}
{\includegraphics{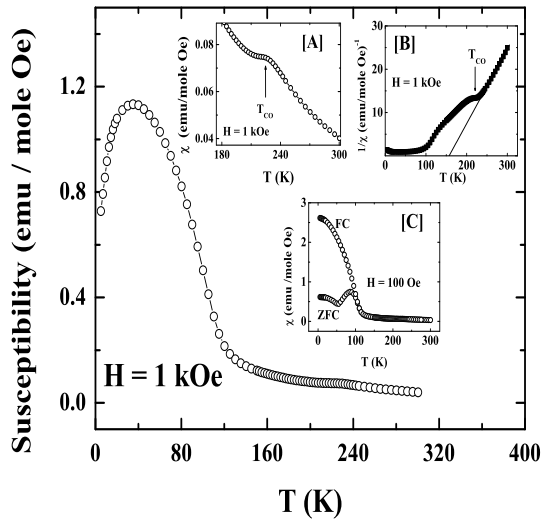}}
\caption{The temperature dependence of zero field cooled dc susceptibility for
nanocrystalline sample of Pr$_{0.65}$Ca$_{0.35}$MnO$_{3}$. The measurements 
have
 been performed in the presence of $1$ kOe magnetic field. Inset:[A] 
The temperature dependence of susceptibility in high temperature region, CO
 transition temperature is indicated by arrow.[B] Inverse zero field
 cooled dc susceptibility
  at $1$ kOe magnetic field has been plotted as a function of temperature. [C] 
The temperature dependence of zero field cooled and field cooled dc 
susceptibility in the
 presence of $100$ Oe magnetic field.}
\end{figure}

The melting of CO state by magnetic field in the bulk sample is
 related with the CAF structure \cite{prca1}. The melting occurs below the 
temperature of 
the onset of CAF structure (T$_{CA}$) 
\cite{prca1}. The similar feature has been observed for the nanocrystalline
 sample. The temperature, T$_{CA}$ for the nanoparticles almost coincides with
 that obtained for the bulk sample \cite{bulk}. 
 At low temperature (below $\sim$ $30$ K), the susceptibility starts to drop
 down with the decreasing temperature as seen in case of similar
 kind of the bulk sample which
 is believed to be  as a result of a spin reorientation 
transition \cite{lees}.
     
We have also performed the zero field cooled (ZFC) and field cooled (FC)
 DC susceptibility measurement in presence of the $100$ Oe magnetic field 
(inset [C],Fig. 4).
A large bifurcation between FC and ZFC curve has been observed 
below $\sim$ T$_{CA}$.
 The surface layers of the nanoparticles are in disordered magnetic
 state comprising of non-collinear spin arrangement as well as defects, 
vacancies. Due to the existence of this disordered surface layer, the magnetic
 frustration can occur and as a result, cluster glass state may form.
The bifurcation between ZFC and FC susceptibility curve is indication of
 this glassy behavior \cite{jap1,ab1}.

All the measurements consistently indicate that CO transition occurs in
 the nanoparticles of Pr$_{0.65}$Ca$_{0.35}$MnO$_{3}$. The CO
 transition temperature in nanoparticles remains almost same as that of
 the bulk form \cite{bulk} of 
the sample. This result
 directly contradicts with the previous experimental results for nanocrystalline
 Pr$_{0.5}$Sr$_{0.5}$MnO$_{3}$ \cite{ab1} and Nd$_{0.5}$Sr$_{0.5}$MnO$_{3}$
 \cite{ab2}.
 
For nanoparticles of 
Pr$_{0.5}$Sr$_{0.5}$MnO$_{3}$, ferromagnetic
 transition temperature (T$_{C}$) remains same as the bulk sample
 and CO transition is not observed down to $4$ K\cite{ab1}. In case of
 Nd$_{0.5}$Sr$_{0.5}$MnO$_{3}$, T$_{C}$ decreases in comparison with the
 bulk. However CO transition is still invisible down to $2$ K \cite{ab2}.
The ferromagnetic transition is governed by DE interaction. There is no
 evidence of strengthening of ferromagnetic DE interaction 
(as reflected in the value of T$_{C}$) due to the reduction of particle size 
for both the two samples. In spite of this fact, CO transition is 
hindered \cite{ab1}.    
 The reduction of particle size has hardly any effect on the on site coulomb
 interaction \cite{brink2}.
Thus, there may be other factor which plays dominant role in CO in
 case of  nanoparticles. 
Charge Order transition has all the essential signatures of martensitic 
transformation \cite{podzorov}. During martensitic transformation, due to the
 nucleation of new crystal structure with in the parent crystal,  lattice
 misfit arises and it gives rise to a strain known as 
martensitic strain \cite{fine}. This strain has to be accommodated by the
 system in order to occur the transformation. The development of the 
martensitic strain depends on the crystal structure of the parent phase and 
the martensitic phase \cite{fine}. 
It is not easy for nanoparticles to 
accommodate the martensitic strain \cite{podzorov,fine}.
   
  During the CO transition of
Pr$_{0.5}$Sr$_{0.5}$MnO$_{3}$, the high temperature  tetragonal crystal 
structure is changed to CO phase of monoclinic crystal structure 
\cite{prsrstruc2,prsrrapid}. For Nd$_{0.5}$Sr$_{0.5}$MnO$_{3}$,CO transition is 
 accompanied by high temperature orthorhombic crystal structure
 to the monoclinic crystal structure \cite{rao} . 
In both the  cases, CO is associated with the
transformation from the 
higher symmetric crystal structure to relatively lower symmetric crystal 
structure. On the other hand Pr$_{0.65}$Ca$_{0.35}$MnO$_{3}$ 
transforms from orthorhombic structure to relatively higher symmetric 
tetragonal (pseudo) crystal structure during CO transition 
\cite{structure1,lees}. 
It seems that, due to the transition from lower to relatively higher crystal 
symmetry, the 
martensitic strain developed during CO transition in  
Pr$_{0.65}$Ca$_{0.35}$MnO$_{3}$ is smaller
 in comparison with Pr$_{0.5}$Sr$_{0.5}$MnO$_{3}$ and  
Nd$_{0.5}$Sr$_{0.5}$MnO$_{3}$. As a result, the 
nanoparticles of 
Pr$_{0.65}$Ca$_{0.35}$MnO$_{3}$ can accommodate the strain which is
 rather difficult for other two  cases. The charge order transition is possible for 
the nanoparticles of Pr$_{0.65}$Ca$_{0.35}$MnO$_{3}$. However,
 the formation of CO state is largely affected in case
 of nanocrystalline Pr$_{0.5}$Sr$_{0.5}$MnO$_{3}$ and  
Nd$_{0.5}$Sr$_{0.5}$MnO$_{3}$. 
The observed behaviors of the nanoparticles of Pr$_{0.65}$Ca$_{0.35}$MnO$_{3}$
 in the present case and Pr$_{0.5}$Sr$_{0.5}$MnO$_{3}$ and  
Nd$_{0.5}$Sr$_{0.5}$MnO$_{3}$ in previous
 cases \cite{ab1,ab2} show that the martensitic like character of transition 
itself plays the dominant role in CO transition.

\section{Summary}
We have prepared and characterized nanoparticles of 
Pr$_{0.65}$Ca$_{0.35}$MnO$_{3}$ with average particle size $\sim$ $40$ nm.
The transport, electron diffraction, heat capacity and magnetization studies
 have been carried out on the sample. All the measurements indicate that the
charge order transition occurs for the sample at almost same temperature
as that of
 the bulk form of the sample. 
The present experimental result lead to the conclusion that the 
martensitic nature of the transition is the key factor for the 
occurrence of CO in nanocrystalline sample of rare-earth manganites.

\section{Acknowledgments}
Authors would like to thank Pulak Ray of Saha Institute of Nuclear Physics, 
Kolkata for providing TEM facility
 and  P. V. Satyam and Jay Ghatak of Institute of Physics, Bhubaneswar, for 
low temperature electron diffraction measurements.


\begin{thebibliography} {abc}

\bibitem{collosal}Colossal Magnetoresistive Oxides, Ed. Y. Tokura 
(Gordon and Breach Science Publishers)

\bibitem{nagaev} A. P. Ramirez, J. Phys: Condens. Matter  {\bf 9}, 8171 (1997).






\bibitem{brink1} J.vandenBrink and D. Khomskii, Phys. Rev. Lett {\bf 82},
1016 (1999).

\bibitem{brink2} J.vandenBrink, G. Khaliullin and D. Khomskii,
Phys. Rev. Lett {\bf 83},5118 (1999).

\bibitem{podzorov} V. Podzorov, B. G. Kim, V. Kiryukhin, M. E. Gershenson
and S-W. Cheong, Phys. Rev. B, {\bf 64}, 140406(R) (2001).
\bibitem{mar1} M. Uhera, S.-W. cheong, Europhys. Lett., {\bf 52}, 674 (2000).
\bibitem{mar2} E. Fertman, D. Sheptyakov, A. Beznosov, V. Desnenko
 and D. Khalyavin, J. Magn. Magn. Mater., {\bf 293}, 787 (2005).

\bibitem{urushihara} A. Urushibara, Y. Moritomo, T. Arima, A. Asamitsu,
G. Kido and Y. Tokura, Phys. Rev. B, {\bf 51}, 14103 (1995).

\bibitem{bulk} V. Dediu, C. Ferdeghini, F. C. Matacotta, P. Nozar and
 G. Ruani, Phys. Rev. Lett., {\bf 84}, 4489 (2000).




\bibitem{physica} Osami Yanagisawa, M. Izumi, W. Z. Hu, K. H. Huang, 
K. Nakamishi and H. Nojima, Physica B, {\bf 271}, 235 (1999).

\bibitem{scherrer} A. J. C. Wilson, Proc. Phys. Soc. london., {\bf 80}, 286, 
1962.
\bibitem{warrer} B.E. Warren, X-ray Diffraction (Addison-Wesley, New york,1969).
\bibitem{cryst} T. R. anantharaman and J. W. Christian, Acta Crystallogr., 
 {\bf 9}, 479 (1956). 

\bibitem{prca1} Y. Tomioka, A. Asamitsu, H. Kuwahara, Y. Moritomo and
 Y. Tokura, Phys. Rev. B, {\bf 53}, R1689 (1996).
\bibitem{chen} C. H. Chen, S.-W. Cheong, Phys. Rev. Lett., {\bf 76}, 4042 
(1996).
\bibitem{edprl} A. P. Ramirez, P. Schiffer, S-W. Cheong, C. H. Chen, W. Bao,
 T. T. M. Palstra, P. L. Gammel, D. J. Bishop and B. Zegarski,
 Phys. Rev. Lett., {\bf 76}, 3188 (1996).
\bibitem{liu} K. Liu, X. W. Wu, K. H. Ahn, T. Sulchek, C. L. Chien
 and John Q. Xiao, Phys. Rev. B, {\bf 54}, 3007 (1996).

\bibitem{lees} M. R. Lees, J. Barratt, G. Balakrishnan, D.M. Paul and
 M. Yethiraj, Phys. Rev. B, {\bf 52}, R14303 (1995).
\bibitem{jap1} Zhi-Hong Wang, Tian-Hao Ji, Yi-Qian Wang, Xin Chen, Run-Wei Li,
 Jian-Wang Cai, Ji-Rong Sun and Bao-Gen Shen, J. Appl. phys., {\bf 87},
 5582 (2000).
\bibitem {ab1} Anis Biswas, I. Das and Chandan Majumdar, J. Appl. Phys.,
 {\bf 98}, 124310 (2005).
\bibitem{ab2} Anis Biswas and I. Das (private communication).

\bibitem{fine} 'Martensitic Transition' Ed. by M. Fine, M. Meshii and
 C. M. Wayman( Academic,New York,1978).



\bibitem{prsrstruc2} A. Llobet, J.L. Garcia-Munoz, C. Frontera and
 C. Ritter,Phys. Rev. B, {\bf 60}, R9889 (1999)
\bibitem{prsrrapid} R. Kajimoto, H. Yoshizawa, Y. Tomioka and Y. Tokura,
Phys. Rev. B, {\bf 66}, 180402(R) (2002).

\bibitem{rao} C. Ritter, R. Mahendiran, M.R. Ibarra, L. Morellon, A. maignan,
B. Raveau and C. N. R. Rao, Phys. ReV. B, {\bf 61}, 9229 (R) (2000).
\bibitem{structure1} Z. jirak,S. Krupicka, V. Nekvasil, E. Pollert
, G. Villeneuve and F. Zounova, J.Magn. Magn. Mater., {\bf 15-18}, 519 (1980).

%

                                                                               
                                                                             



\end{thebibliography}
\end{document}